\documentclass[letterpaper,11pt]{article}
\pdfoutput=1
\usepackage{jcappub}
\usepackage{etoolbox}
 \makeatletter
    \patchcmd{\maketitle}{\@fpheader}{}{}{}
    \makeatother
\usepackage{threeparttable} 
\usepackage[T1]{fontenc}
\usepackage{color}
\usepackage{epstopdf}
\usepackage{natbib}
\usepackage{mathrsfs}
\usepackage{subcaption}
\usepackage{tikz-cd}
\usepackage{mdwlist}
\usepackage{dsfont}
\usepackage{natbib}
\usepackage{amssymb}
\usepackage{amsthm}
\usepackage{graphicx}
\usepackage{amsmath}
\usepackage{url}
\usepackage{bm}
\usepackage{epstopdf}
\usepackage{color}
\usepackage{enumitem}
\usepackage{dsfont}
\usepackage{bbm}
\usepackage{mathrsfs}
\usepackage[utf8]{inputenc}

\newcommand{\di}{\partial}
\newcommand{\ident}{\hat{\mathbbm{1}}}

\newcommand{\Pl}{\text{Pl}}
\newcommand{\LL}[1]{\hat{\mathcal{L}}_{#1}}
\newcommand{\GG}[1]{\hat{\mathcal{L}}^{-1}_{#1}}
\renewcommand{\L}{\hat{\mathcal{L}}}

\newcommand{\mink}{\mathbb{H}^{1,3}}
\newcommand{\G}{\mathcal{G}}

\def\be {\begin{equation}}

\def\ee  {\end{equation}}

\def\bea {\begin{eqnarray}}

\def\eea {\end{eqnarray}}

\def\nn {\nonumber}

\begin{document}

\title{The initial value problem for gravitational waves in conformal gravity}

\author{Sanjeev S.\ Seahra}

\affiliation{Department of Mathematics and Statistics, University of New Brunswick \\ Fredericton, NB, Canada E3B 5A3}

\emailAdd{sseahra@unb.ca}

\abstract{In certain models of conformal gravity, the propagation of gravitational waves is governed by a fourth order scalar partial differential equation.  We study the initial value problem for a generalization of this equation, and derive a Kirchhoff-like explicit solution in terms of the field and its first three time derivatives evaluated on an initial hypersurface, as well as second order spatial derivatives of the initial data.  In the conformal gravity case, we establish that if the initial data is featureless on scales smaller than the length scale of conformal symmetry breaking, then we recover the ordinary behaviour of gravitational waves in general relativity.  We also confirm that effective weak field gravitational force exerted by a static spherical body in such models becomes constant on small scales; i.e.\ conformal gravity is effectively 2-dimensional at high energies.}


\date{\today}

\maketitle

\section{Introduction}

Our current best understanding of cosmology and gravitation has some well-known undesirable features:  Einstein's general relativity is inconsistent with observations of galactic dynamics, gravitational lensing, and the expansion history of the universe without assuming the existence of dark matter that is virtually impossible to detect non-gravitationally.  The observed late time acceleration of the universe is best modelled by introducing a cosmological constant into the gravitational action that takes on a value hundreds of orders of magnitude smaller than a na\"ive first estimate.  Finally, attempts to apply straightforward techniques from quantum field theory to quantize gravitational fluctuations around flat space lead to incurable divergences in physical quantities.

Many authors have postulated that the root of some, if not all, of these problems is that Einstein's theory is not the true theory of gravitation \cite[reviews]{Clifton:2011jh,Ishak:2018his}.  However, exchanging general relativity for some modified gravity model is not trivially done:  General relativity is extremely accurate in describing astrophysical phenomena on sub-galactic scales.  The theory passes very precise Solar System tests, accurately predicts the spin-down of binary pulsars, and has recently been shown to be consistent with observation of gravitational wave signals from binary black hole and neutron star mergers  \cite{Abbott:2016blz,Yunes:2013dva,TheLIGOScientific:2016src,Yunes:2016jcc,TheLIGOScientific:2016pea}.  It is therefore imperative that any proposed modified theory of gravity be tested against astrophysical observations that are entirely consistent with general relativity to within experimental errors \cite[review]{Berti:2015itd}.

One theoretically appealing approach to modifying gravity involves assuming there exists a fundamental local conformal (Weyl) symmetry that is broken at low energy \cite{weyl,bach,Mannheim:1989jh,Mannheim:2007ug,Mannheim:1992tr,Mannheim:1999bu,Mannheim:2011ds}.  Some of the diverse motivations for such an approach are outlined in \cite{Hooft:2014daa}: but one of the most compelling is that such theories are power-counting renormalizable \cite{Stelle:1977ry}.  However, there are several important negative features of such models.  Generally, they are fourth-order theories of gravity, and are hence typically plagued by perturbative ghosts.  Some authors have claimed the ghost issue can be solved via various mechanisms \cite{Mannheim:2000ka,Bender:2007nj,Bender:2007wu,Bender:2008gh,Bender:2008vh,Mannheim:2009zj,Mannheim:2015hto,Mannheim:2015tlu}.  Also, since the world that we live is clearly not conformally invariant, such models require a mechanism to spontaneously break conformal symmetry at low energies.

In this paper, we study the classical dynamics of gravitational waves model featuring a dynamically broken conformal symmetry \cite{holscher,Gegenberg:2017bms,Caprini:2018oqe,Holscher:2019swu}.  Specifically, we are interested in the complete solution of the initial value problem for vacuum gravitational waves in conformal gravity in a certain gauge.  We will actually solve a slightly more general problem: that of a fourth order partial differential equation that can be interpreted as the wave equation obtained when two ordinary Klein-Gordan differential operators act in succession to annihilate a scalar field.  We will obtain the generalized Kirchhoff formula solution for the scalar field in terms its zeroth, first, second and third time derivative on an initial hypersurface.  For the conformal gravity case, we establish the circumstances under which we recover the ordinary propagation of gravitational waves in general relativity: namely, if the initial data and source are featureless on scales smaller than the scale of conformal symmetry breaking, then then deviations from general relativity are negligible.

The plan of the paper is as follows:  In section \ref{sec:conformal}, we review the equations of motion for gravitational waves in conformal gravity models.  In section \ref{sec:Green's}, we introduce a generalization of the conformal gravitational wave equation of motion and derive the associated retarded Green's function.  In section \ref{sec:Kirchhoff}, we use this Green's function to solve the initial value problem in general, and then specifically for the conformal gravity case.  Section \ref{sec:discussion} is reserved for a discussion of our results.

\section{Gravitational waves in conformal gravity}\label{sec:conformal}

Our starting point is the effective field equations of conformal gravity as presented in Refs.~\cite{Gegenberg:2017bms,Caprini:2018oqe,Holscher:2019swu}\footnote{Note that we follow the Misner-Thorne-Wheeler sign conventions, so (\ref{eq:field eqn}) differs from the field equations presented in \cite{Caprini:2018oqe,Holscher:2019swu} by the sign of the Einstein tensor.}:
\begin{equation}\label{eq:field eqn} 	
\epsilon G_{\alpha\beta} + \Lambda g_{\alpha\beta} + 2 M_{\star}^{-2}
B_{\alpha\beta} = {M_\Pl^{-2}} T_{\alpha\beta}, \end{equation} 
where $M_{\star}$ is a mass scale above which exotic physics effects become important; $M_{\Pl} = (8\pi G)^{-1/2}$ is the reduced Planck mass; $G_{\alpha\beta}$ and $T_{\alpha\beta}$ are the Einstein and stress-energy tensors as usual; $\epsilon = \pm 1$; and $B_{\alpha\beta}$ is the conformally invariant Bach tensor defined by: 
\begin{subequations}
\begin{align} 
B_{\mu\nu} & = -\nabla^{\alpha} \nabla_{\alpha} S_{\mu\nu} + \nabla^{\alpha} \nabla_{\mu} S_{\alpha\nu} + C_{\mu\alpha\nu\beta} S^{\alpha\beta}, \\ S_{\alpha\beta} & = \tfrac{1}{2}(R_{\alpha\beta} - \tfrac{1}{6} R g_{\alpha\beta}), 
\end{align} \end{subequations} 
where $C_{\alpha\beta\gamma\delta}$ is the Weyl tensor.  

The field equation (\ref{eq:field eqn}) can be derived \cite{Mannheim:1989jh,Caprini:2018oqe,Holscher:2019swu} from a conformally invariant action of the form 
\begin{equation}\label{eq:action 2} 	
S =  -\frac{M_\Pl^{2}}{M_{\star}^{2}} \int d^{4}x \sqrt{-g} \, C^2 + S_{\Psi} + S_\text{m},
 \end{equation}
where $S_{\Psi}$ is the action of a conformally coupled scalar field $\Psi$, where $C^{2} = C_{\alpha\beta\gamma\delta}C^{\alpha\beta\gamma\delta}$, and $S_\text{m}$ is the matter action.  The vacuum expectation value of this scalar field is $\Psi = \Psi_{0} =$ constant.   This solution spontaneously breaks conformal symmetry, and when it is substituted back into the action we obtain
\begin{equation}\label{eq:action 1} 	
S =  \frac{M_\Pl^{2}}{2} \int d^{4}x \sqrt{-g} \, \left(\epsilon R - 2\Lambda - \frac{2}{M_{\star}^{2}} C^2 \right)+ S_\text{m},
 \end{equation} 
where the parameter $\epsilon = \pm 1$ depends on the details of the conformal symmetry breaking mechanism; in the original work of \citet{Mannheim:1989jh}, the choice $\epsilon = -1$ was made.  

The main motivation to study actions of the form (\ref{eq:action 1}) is that at high energies the $C^{2}$ term (which involves 4 derivatives of the metric) will dominate over the Einstein-Hilbert term (which has 2 derivatives).  This results in a model which is conformally invariant at high energy and potentially renormalizable, but reduces to ordinary general relativity at low energy.  The mass scale $M_{\star}$ controls the energy scale of the transition between these two behaviours.  Note that in order to recover general relativity at low energy, one need to choose $\epsilon = +1$, which is the opposite of the original choice of \citet{Mannheim:1989jh}.  Despite this problem with the $\epsilon = -1$ case, we include it in our work below for completeness.

We will neglect the cosmological constant $\Lambda$ below (the dynamics of gravitational waves including finite $\Lambda$ effects were considered in \cite{Gegenberg:2017bms}).  Note that the Bach tensor involves fourth order derivatives of the metric, hence this is a higher derivative theory of gravity.  

If (\ref{eq:field eqn}) is linearized about flat space such that $g_{\alpha\beta}=\eta_{\alpha\beta}+h_{\alpha\beta}$, the equation of motion for metric perturbations $h_{\alpha\beta}$ in the so-called Teyssandier gauge \cite{Teyssandier_1989} is \cite{Gegenberg:2017bms,Caprini:2018oqe,Holscher:2019swu}:
\begin{subequations}\label{eq:GW EOM}
\begin{gather}
	M_{\star}^{-2} (\Box - \epsilon M_{\star}^{2}) \Box h_{\alpha\beta} = 2M_{\Pl}^{-2} \tilde{T}_{\alpha\beta}, \\
	\tilde{T}_{\alpha\beta} = T_{\alpha\beta} - \tfrac{1}{2} \eta_{\alpha\beta} T + \tfrac{1}{6} \epsilon M_{\star}^{-2} \eta_{\alpha\beta} \Box T.
\end{gather}	
\end{subequations}
Here, $\Box = -\di_{t}^{2} + \nabla^{2}$ and $T_{\alpha\beta}$ is the stress-energy tensor of matter, which is assumed to be the same order as $h_{\alpha\beta}$.  We see that if $M_{\star} \rightarrow \infty$, we recover the ordinary equation of motion for gravitational waves in general relativity if $\epsilon = +1$.

Because the differential operator has a trivial index structure, the initial value problem associated with (\ref{eq:GW EOM}) is formally equivalent to solving
\begin{equation}\label{eq:GW EOM scalar}
	(\Box - m^{2})\Box\phi(x) = \alpha F(x),
\end{equation}
for the unknown function $\phi: \mink \rightarrow \mathbb{R}$ representing the components of $h_{\alpha\beta}$.\footnote{$\mink$ refers to 4-dimensional Minkowski space.}  Here, $m^{2} = \epsilon M_{\star}^{2}$ is a possibly imaginary mass parameter, $F: \mink \rightarrow \mathbb{R}$ is a given source function representing the components of $T_{\alpha\beta}$, and $\alpha = 2M_{\star}^{2}/M_{\Pl}^{2}$ is a coupling constant.  Therefore, in the subsequent sections we will concentrate on a generalization of this type of scalar field equation.

\section{Green's functions for the generalized fourth order wave equation}\label{sec:Green's}

In this section, we develop the Green's function for equations like (\ref{eq:GW EOM scalar}).  Because it takes very little additional effort, we will work with a slight generalization of (\ref{eq:GW EOM scalar}):
\begin{equation}\label{eq:PDE 1}
	(\Box - m_{1}^{2})(\Box - m_{2}^{2})\phi(x) = \alpha F(x), \quad x = (t,\mathbf{x}) \in \mathbb{H}^{1,3}.
\end{equation}
The two masses $m_{1}$ and $m_{2}$ may be equal to one another.  We will be concerned with the initial value (Cauchy) problem for $\phi$.  Therefore, we specify initial data  on a $t=t_{0} = $ constant hypersurface:
\begin{equation}\label{eq:ICs}
	\phi\big|_{t=t_{0}} = \Phi_{0}(\mathbf{x}), \quad \di_{t}\phi\big|_{t=t_{0}} = \Phi_{1}(\mathbf{x}), \quad \di_{t}^{2}\phi\big|_{t=t_{0}} = \Phi_{2}(\mathbf{x}), \quad \di_{t}^{3}\phi\big|_{t=t_{0}} = \Phi_{3}(\mathbf{x}).
\end{equation}

We will find in useful to re-write (\ref{eq:PDE 1}) in terms of differential operators:
\begin{gather}\label{eq:PDE 2}
	\LL{0} \phi(x) = \alpha F(x), \\ \label{eq:operator defs}
	 \LL{0} = \LL{1}\LL{2}, \quad \LL{1} = \Box - m_{1}^{2}, \quad \LL{2} = \Box - m_{2}^{2}.
\end{gather}
We note that the two second order operators obviously commute $[\LL{1},\LL{2}] = 0$ and satisfy
\begin{equation}\label{eq:differ by const}
	\LL{1} - \LL{2} = (m_{2}^{2} - m_{1}^{2}) \ident.
\end{equation}
The identity $\ident$ is defined by
\begin{equation}
	\ident f(x) = \int d^{4}x \, \delta^{(4)}(x-x') f(x') = f(x),
\end{equation}
where here and below $f : \mathbb{H}^{1,3} \to \mathbb{R}$ is an arbitrary test function, and $\delta^{(4)}(x-x')$ is the 4-dimensional Dirac $\delta$-function (distribution).

Associated with any given $n^\text{th}$ order differential operator $\L : C^{4}(\mink) \to C^{4-n}(\mink)$ with $n \le 4$, one can define a right inverse operator $\GG{}{} : C^{4-n}(\mink) \to C^{4}(\mink)$ such that
\begin{equation}
	\L \GG{}{} = \ident.
\end{equation}
As usual, we take the action of $\GG{}{}$ on a test function to be a convolution with a Green's function $G$:
\begin{equation}
	\GG{}{} f(x) = \int d^{4}x' \, G(x-x')f(x'), 
	\quad \L G(x-x') = \delta^{(4)}(x-x').
\end{equation}
Note that it is not true that $\GG{}\L f = f$ unless $f$ vanishes at infinity.

In general, $\GG{}{}$ is not uniquely defined unless one imposes boundary conditions on $G$.  In this work, we will concentrate exclusively on retarded boundary conditions, which state that $G(x-x') = 0$ if $x \notin J^{+}(x')$, where $J^{+}(x')$ is the causal future $x'$:
\begin{equation}
	J^{+}(x') = \left\{ x = (t,\mathbf{x}) \in \mink \mid (t-t')^{2} \ge |\mathbf{x}-\mathbf{x}'|^{2}, t > t' \right\}.
\end{equation}
We will denote the inverse of the $\LL{i}$ operators defined in (\ref{eq:operator defs}) by $\GG{i}{}$ and the associated Green's functions by $G_{i}$.

Now, since $ \LL{0} = \LL{1}\LL{2} = \LL{2}\LL{1}$, it follows that
\begin{multline}
	\GG{0} = \GG{2}\GG{1} =\GG{1}\GG{2} \,\, \Rightarrow \\ G_{0}(x-x') = \int d^{4}x'' G_{2}(x-x'') G_{1}(x''-x') = \int d^{4}x'' G_{1}(x-x'') G_{2}(x''-x').
\end{multline}
These imply the identities
\begin{equation}\label{eq:identities}
	\LL{1} G_{0}(x-x') = G_{2}(x-x'), \quad \LL{2} G_{0}(x-x') = G_{1}(x-x').
\end{equation}
Note that the retarded $G_{1}$ and $G_{2}$ Green's functions are known explicitly \cite{Poisson:2011nh}
\begin{equation}\label{eq:known Green's}
	G_{i}(x-x') = \frac{\Theta(t-t')}{4\pi} \left[ -\delta(\sigma) + \frac{\Theta(-\sigma) m_{i} J_{1}(m_{i}\sqrt{-2\sigma})  }{\sqrt{-2\sigma}} \right], \quad i = 1,2,
\end{equation}
where $\Theta$ is the Heaviside function, $J_{1}$ is the Bessel function of the first kind of order 1, and
\begin{equation}
	\sigma = \sigma(x-x') = - \frac{(t-t')^{2} - | \mathbf{x}-\mathbf{x}' |^{2} }{2}
\end{equation}
is Synge's world function, which is a Lorentz invariant.  

We now derive a formula for the Green's function $G_{0}$ satisfying  $\LL{0} G_{0}(x-x') = \delta^{(4)}(x-x')$ with retarded boundary conditions when $m_{1} \ne m_{2}$:
\begin{align}
	G_{0}(x-x')  = \frac{\Theta(t-t')\Theta(-\sigma)}{4\pi\sqrt{-2\sigma}} \left[ \frac{ m_{1} J_{1}(m_{1}\sqrt{-2\sigma}) - m_{2} J_{1}(m_{2}\sqrt{-2\sigma}) }{m_{1}^{2} -m_{2}^{2}} \right].\label{eq:Green's formula}
\end{align}
To show this, consider
\begin{align}
	\nn \L_{0} [ (m_{1}^{2}-m_{2}^{2})^{-1} (\GG{1}{}-\GG{2}{}) ] & =  (m_{1}^{2}-m_{2}^{2})^{-1} (\LL{2}\LL{1}\GG{1}{} - \LL{1}\LL{2}\GG{2}{}) \\ \nn & = (m_{1}^{2}-m_{2}^{2})^{-1} (\LL{2} - \LL{1}) \\   & = \ident; \label{eq:proof}
\end{align}
i.e., $\GG{0}{} =  (m_{1}^{2}-m_{2}^{2})^{-1} (\GG{1}{}-\GG{2}{})$.  From this, it follows that
\begin{equation}
	\int d^{4}x \left\{ \LL{0} \left[ \frac{G_{1}(x-x')-G_{2}(x-x')}{m_{1}^{2} - m_{2}^{2}} \right] - \delta^{(4)}(x-x') \right\} f(x') =0,
\end{equation}
where $f$ is an arbitrary test function, and $G_{1.2}$ are Green's functions of $\LL{1,2}$.  Since $f$ is arbitrary, the quantity in curly brackets must be zero.  That is,
\begin{equation}\label{eq:G0 result}
	G_{0}(x-x') = \frac{G_{1}(x-x')-G_{2}(x-x')}{m_{1}^{2} - m_{2}^{2}}
\end{equation}
is a solution of $\LL{0} G_{0}(x-x') = \delta^{(4)}(x-x')$.  Furthermore, if we take $G_{1,2}$ to be the retarded Green's functions given in (\ref{eq:known Green's}), $G_{0}$ satisfies retarded boundary conditions and we obtain (\ref{eq:Green's formula}).  In Appendix \ref{sec:direct}, we provide an alternate derivation of (\ref{eq:Green's formula}).  Note that for the conformal gravity case, we can take $m_{1}^{2} = 0$ and $m_{2}^{2} = \epsilon M_{\star}^{2}$, which leads to
\begin{equation}\label{eq:G0 result}
	G_{0}(x-x') = \frac{G_{2}(x-x')-G_{1}(x-x')}{\epsilon M_{\star}^{2}}.
\end{equation}

To obtain a guess for the Green's function in the degenerate $m_{1} = m_{2} = m$ case, we set
\begin{equation}
	m_{1} = m, \quad m_{2} = m + \varepsilon,
\end{equation}
in (\ref{eq:Green's formula}) and take the limit $\varepsilon \rightarrow 0$.  We obtain:
\begin{align}\label{eq:degenerate Greens}
	G_{0}(x-x')  = \frac{\Theta(t-t')\Theta(-\sigma)J_{0}(m\sqrt{-2\sigma})}{8\pi}.
\end{align}
We confirm this formula in Appendix \ref{sec:direct} by solving $\LL{0} G_{0}(x-x') = \delta^{(4)}(x-x')$ directly.  Finally, the $m \rightarrow 0$ limit of (\ref{eq:degenerate Greens}) yields
\begin{align}\label{eq:degenerate Greens}
	G_{0}(x-x')  = \frac{\Theta(t-t')\Theta(-\sigma)}{8\pi}.
\end{align}
This is the retarded Green's function of the ``box-squared'' operator $\Box^{2} = (-\di_{t}^{2} + \nabla^{2})^{2}$.  It has been previously calculated by \citet{holscher}.

Before moving on, we comment that the derivation (\ref{eq:proof}) holds for equally well for any two differential operators that differ by a constant; i.e., that satisfy (\ref{eq:differ by const}).  So, we could write
\begin{equation}\label{eq:generalized L's}
	\L_{1} = \hat{\mathcal{Q}} - m_{1}^{2}, \quad  \L_{2} = \hat{\mathcal{Q}} - m_{2}^{2},
\end{equation}
where $\hat{\mathcal{Q}}$ is a $n^\text{th}$ order differential operator.  Then, (\ref{eq:G0 result}) will hold with $G_{1}$ and $G_{2}$ being the Green's functions of the operators in equation (\ref{eq:generalized L's}).\footnote{Note that if $\L_{1,2}$ are $n^\text{th}$ order operators, then $\L_{0} : C^{2n}(\mink) \to C^{0}(\mink)$.  Also, in this case $m_{1}$ and $m_{2}$ do not necessarily have dimensions of mass.}

\section{Initial value problem for the generalized fourth order wave equation}\label{sec:Kirchhoff}

\subsection{Generalized Kirchhoff's formula}

We now turn our attention to the initial value problem for the PDE (\ref{eq:PDE 1}) with initial data given by (\ref{eq:ICs}).  Consider an arbitrary region of spacetime $\Omega$ with boundary $\di\Omega$.  Also, let $n^{\alpha}$ be a normal vector to $\di\Omega$ that is outward pointing when $\di\Omega$ is timelike and inward pointing when $\di\Omega$ is spacelike.  Using the divergence theorem, we have this modified version of Green's second identity:
\begin{equation}
	\int_{\Omega} d^{4}x \, \LL{i} \psi = \int_{\Omega}  d^{4}x \, \psi \LL{i} \phi + \int_{\di\Omega} dS \, n^{\alpha} (\phi \overleftrightarrow{\di_{\alpha}} \psi), \quad i = 1,2,
\end{equation}
where $\psi$ and $\phi$ can be taken to be either scalar functions or distributions (such that the integrals appearing above are well-defined).  Also, $dS^{\alpha} = n^{\alpha} dS$ is the directed surface element on $\di\Omega$. If we take $i=1$, $\phi$ to be a solution of (\ref{eq:PDE 2}), and $\psi = \LL{2}'G_{0}(x-x')$ with $x \in \Omega$, we get
\begin{equation}
	\phi(x) = \int\limits_{x'\in\Omega}  d^{4}x \, [\LL{1}'\phi(x')] [\LL{2}'G_{0}(x-x')] + \int\limits_{x'\in\di\Omega}dS \, n^{\alpha} [ \phi(x') \overleftrightarrow{\di'_{\alpha}} \LL{2}' G_{0}(x-x')].
\end{equation} 
Here, the prime on the differential operators is meant to indicate differentiation with respect to $x'$.  Applying Green's identity again to the first integral on the righthand side, we get
\begin{equation}
	\phi(x) =  \!\!\!\! \int\limits_{x'\in\Omega}  d^{4}x \,
	G_{0}(x-x')\alpha F(x') +  \!\!\!\!  \int\limits_{x'\in\di\Omega} dS \, n^{\alpha} [ \phi(x') \overleftrightarrow{\di'_{\alpha}} G_{1}(x-x') + (\LL{1}'\phi(x')) \overleftrightarrow{\di'_{\alpha}} G_{0}(x-x')],
\end{equation}
where we have made use of the fact that $\LL{2}'G_{0}(x-x') = G_{1}(x-x')$.  Note that we can derive an equivalent formula by repeating this derivation with $\psi = \LL{1}'G_{0}(x-x')$:
\begin{equation}
	\phi(x) = \!\! \int\limits_{x'\in\Omega}  d^{4}x \,
	G_{0}(x-x')\alpha F(x') + \!\!\!\! \int\limits_{x'\in\di\Omega} dS \, n^{\alpha} [ \phi(x') \overleftrightarrow{\di'_{\alpha}} G_{2}(x-x') + (\LL{2}'\phi(x')) \overleftrightarrow{\di'_{\alpha}} G_{0}(x-x')],
\end{equation}
Subtracting the two expressions yields:
\begin{equation}
	0 =  \int\limits_{x'\in\di\Omega} dS \, n^{\alpha} [\phi(x') \overleftrightarrow{\di'_{\alpha}} [G_{1}(x-x') -  G_{2}(x-x')  - (m_{1}^{2}-m_{2}^{2}) G_{0}(x-x')].
\end{equation}
Since this holds for arbitrary regions $\Omega$, we can conclude that the quantity in square brackets is zero, reproducing equation (\ref{eq:G0 result}).

We can now specialize to the geometry depicted in figure \ref{fig:figure}.  Making note of the fact that $n^{\alpha} \di_{\alpha}' = \di_{t'}$ and that all Green's functions depend on $\tau = t-t'$, we obtain
\begin{equation}
	\phi(x) = \phi_{F}(x) + \phi_{1}(x) + \phi_{2}(x),
\end{equation}
where
\begin{subequations}\label{eq:phi components}
\begin{align}
	\phi_{F}(x) = &\int\limits_{x'\in\Omega}  d^{4}x \, G_{0}(x-x')  \alpha F(x'), \\
	\nonumber \phi_{1}(x) = & - \frac{\di}{\di\tau} \int\limits_{x'\in\di\Omega} dS \, G_{1}(x-x') \phi(x') - \int\limits_{x'\in\di\Omega}  dS \,G_{1}(x-x')\di_{t'} \phi(x') \\ = & 
	- \frac{\di}{\di\tau} \int\limits_{x'\in\di\Omega}  dS \,G_{1}(x-x') \Phi_{0}(\mathbf{x}') - \int\limits_{x'\in\di\Omega}  dS \,G_{1}(x-x') \Phi_{1}(\mathbf{x}')  , \\ 
	\phi_{2}(x) = & - \frac{\di}{\di\tau} \int\limits_{x'\in\di\Omega}dS \, G_{0}(x-x') \LL{1}'\phi(x')  - \int\limits_{x'\in\di\Omega}dS \, G_{0}(x-x') \LL{1}' \di_{t'} \phi(x') \\ \nonumber = & - \frac{\di}{\di\tau} \int\limits_{x'\in\di\Omega} dS \,G_{0}(x-x') [-\Phi_{2}(\mathbf{x}') + (\nabla_{\mathbf{x}'}^{2}-m_{1}^{2}) \Phi_{0}(\mathbf{x}')]  \\ & - \int\limits_{x'\in\di\Omega}dS \, G_{0}(x-x')  [-\Phi_{3}(\mathbf{x}') + (\nabla_{\mathbf{x}'}^{2}-m_{1}^{2}) \Phi_{1}(\mathbf{x}')].
\end{align}
\end{subequations}
These formulae give the explicit solution of the initial value problem (\ref{eq:PDE 1}) and (\ref{eq:ICs}) in terms of the Green's functions derived in section \ref{sec:Green's}.  From these, the existence and uniqueness of solutions of (\ref{eq:PDE 1}) and (\ref{eq:ICs}) is easily established.  

\begin{figure}
\begin{center}
\begin{tikzpicture}
 \draw[fill=magenta!10] (5,-1.7) -- (2.2,0.9) -- (-1.8,0.9) -- (-4.3,-1.7) 
      -- cycle;

    \shade[left color=blue!5!white,right color=blue!40!white,opacity=0.5]  (-2,0) arc (155:385:2cm and 1cm) -- (0,2) -- cycle;
    
     
     \shade[left color=blue!25!white,right color=blue!60!white,opacity=0.5] (-0.2,-0.42) ellipse (2cm and 1cm);

    \draw[line width=0.3mm] (-2,0) arc (155:385:2cm and 1cm) -- (0,2) -- cycle;
    \draw[dashed] (-2,0) arc (155:25:2cm and 1cm);
    \draw[dashed] (-2.1,-0.3) -- (0,-.3) -- (0,2);
    
    \node[above] at (-2.7,-1.7){$\di\Omega: t'=t_{0}$};
    \node[above] at (0,-1.4){$\di B$};
    \node at (-1,-0.8){$B$};

    \node[above] at (-1,-0.3){$\tau$};
    \node[left] at (0,1.2){$\tau$};
    
    \node[above] at (0,2){$x=(t_{0}+\tau,\mathbf{x})$};

      \draw[black,fill=purple] (0,2) circle (.075 cm);
      
       \draw[dashed] (0,-0.3) -- (2.0,-1.25) node[midway,above] {$R$};
      
        \node[right] at (2.0,-1.25){$x'=(t_{0},\mathbf{x}')$};

      \draw[black,fill=purple] (2.0,-1.25) circle (.075 cm);
    
\end{tikzpicture}
\end{center}
\caption{Spacetime geometry}\label{fig:figure}
\end{figure}

\subsection{Specialization to the conformal gravity case}

We now apply the general results of section \ref{sec:Kirchhoff} to the conformal gravity scenario.  Specifically, we set
\begin{equation}
	m_{1} = 0, \quad m_{2} = \begin{cases} M_{\star}, & \epsilon = +1, \\ iM_{\star}, & \epsilon=-1, \end{cases} \quad \alpha = 2M_{\star}^{2}/M_{\Pl}^{2},
\end{equation}
which yields the Green's functions
\begin{align}\label{eq:conformal Green's}
	G_{0}(x-x') & = \frac{\Theta(\tau)\Theta(\tau-R)}{4\pi M_{\star}\sqrt{\tau^{2}-R^{2}}} \begin{cases} J_{1}(M_{\star}\sqrt{\tau^{2}-R^{2}}), & \epsilon = +1, \\ I_{1}(M_{\star}\sqrt{\tau^{2}-R^{2}}), & \epsilon = -1, \end{cases}\\G_{1}(x-x') & = -\frac{\Theta(\tau)\delta(\tau -R)}{4\pi \tau}.
\end{align}
Here, $I_{1}$ is a modified Bessel function of the first kind.

Plugging these expressions into (\ref{eq:phi components}) above, we obtain the following explicit expression for the solution of the initial value problem when $\epsilon = 1$:
\begin{subequations}
\begin{align}
	\phi(x) = & \phi_{F}(x) + \phi_{1}(x) + \phi_{2}(x), \\ \label{eq:source response}
	\phi_{F}(x) = & \frac{M_{\star}}{2\pi M_{\Pl}^{2} }  \int_{0}^{\tau} d\tau' \int_{0}^{\tau'} dR  \iint \sin\theta\, d\theta \, d\varphi  \,\frac{ R^{2}J_{1}(M_{\star}\sqrt{\tau'^{2}-R^{2}})}{\sqrt{\tau'^{2}-R^{2}}} F(R,\theta,\varphi), \\ \label{eq:ordinary Kirchhoff}
	\phi_{1}(x) = & \frac{1}{4\pi} \frac{\di}{\di\tau} \left( \frac{1}{\tau} \iint_{\di B} \sin\theta\, d\theta \, d\varphi \, \Phi_{0} \right) + \frac{1}{4\pi\tau}\iint_{\di B}\sin\theta\, d\theta \, d\varphi\, \Phi_{1}, \\ \label{eq:higher order correction}
	\phi_{2}(x) = & -\frac{1}{4\pi M_{\star}} \frac{\di}{\di\tau}  \int_{0}^{\tau} dR \iint \sin\theta\, d\theta \, d\varphi   \, \frac{R^{2} J_{1}(M_{\star}\sqrt{\tau^{2}-R^{2}})}{\sqrt{\tau^{2}-R^{2}}}  (-\Phi_{2}+\nabla^{2}\Phi_{0}) \nonumber  \\ & -\frac{1}{4\pi M_{\star}}  \int_{0}^{\tau} dR \iint \sin\theta\, d\theta \, d\varphi   \, \frac{R^{2} J_{1}(M_{\star}\sqrt{\tau^{2}-R^{2}})}{\sqrt{\tau^{2}-R^{2}}}  (-\Phi_{3}+\nabla^{2}\Phi_{1}).
\end{align}
\end{subequations}
The corresponding equations for the $\epsilon = -1$ case are found by exchanging $J_{1}$ for $I_{1}$.

We now comment on the various parts of this solution: First, we note that (\ref{eq:ordinary Kirchhoff}) is just the ordinary Kirchhoff formula for the solution of $\Box \phi =0$ \cite{strauss2007partial}.  It's presence in the full solution of (\ref{eq:GW EOM scalar}) indicates the existence of a massless mode.  

Equation (\ref{eq:source response}) represents the direct response of $\phi$ to the source $F$.  This equation was studied in some detail in \cite{Gegenberg:2017bms}, where it was pointed out that the fact that the Green's function is finite and has support away from the past light cone means that gravitational signals tend to be blurry in conformal gravity.   Let us now consider the limit when $M_{\star}$ is large.  In this case, the argument of the integral will be sharply peaked in neighbourhood of $R= \tau'$.  If we assume that the source does not vary very much in this neighbourhood, then we can take $F(R,\Omega) \approx F(\tau',\Omega)$, which yields:
\begin{align}\nonumber
	\phi_{F}(x) & \approx \frac{1}{2\pi M_{\Pl}^{2} }  \int_{0}^{\tau} {d\tau'}   \iint \sin\theta\, d\theta \, d\varphi   \, F(\tau',\theta,\varphi)  \int_{0}^{\tau'} \, dR \, \frac{ M_{\star}R^{2}J_{1}(M_{\star}\sqrt{\tau'^{2}-R^{2}})}{\sqrt{\tau'^{2}-R^{2}}} \\ \nonumber & = \frac{1}{2\pi M_{\Pl}^{2}  }  \int_{0}^{\tau} {d\tau'}   \iint \sin\theta\, d\theta \, d\varphi   \, F(\tau',\theta,\varphi) \left( \tau' - \frac{\sin M_{\star}\tau'}{M_{\star}} \right) \\ & \approx \frac{1}{2\pi M_{\Pl}^{2}  }  \int_{0}^{\tau} {d\tau'} \tau'  \iint \sin\theta\, d\theta \, d\varphi   \, F(\tau',\theta,\varphi),\label{eq:source equation limit}
\end{align}
 where in the last line, we have assumed that the source only has support on the portion of the lightcone with $M_{\star}\tau' \gg 1$; i.e., the source in not located within a distance of $1/M_{\star}$ of the observation point.  Equation (\ref{eq:source equation limit}) is the usual formula for the solution of $\Box\phi = -2M_{\Pl}^{-2}F$ with ``no incoming radiation'' boundary conditions.
 
It is also interesting to examine (\ref{eq:source response}) for the case of a static and eternal point source located a spatial distance of $R_{0}$ from the observer:
\begin{equation}
	F(R,\theta,\varphi) = \kappa\frac{\delta(R-R_{0})\delta(\theta-\theta_{0})\delta(\varphi-\varphi_{0})}{R^{2}\sin\Theta}.	
\end{equation}
Taking the $\tau\rightarrow \infty$ limit and making use of integral 6.645 in \cite{GR}, we obtain
\begin{equation}
	\phi_{F} = \frac{\kappa (1-e^{-M_{\star}R_{0}})}{2\pi R_{0} M_{\Pl}^{2} }.
\end{equation}
As usual, we can interpret $\phi_{F} = -2U_\text{G}$, where $U_\text{G}$ is the effective weak field gravitational potential outside a static and spherically symmetric body.  Selecting the constant $\kappa$ to match the Newtonian result at large distances, we get
\begin{equation}
	U_\text{G} = - \frac{GM(1-e^{-M_{\star}R_{0}})}{R_{0}}.
\end{equation}
From this, we can calculate the gravitational force $F_\text{G} = - {\di U_\text{G}}/{\di R_{0}}$ exerted by the source on small scales:
\begin{equation}
	F_\text{G} = -\frac{GMM_{\star}^{2}}{2} \left[ 1 - \frac{2}{3} M_{\star}R_{0} + \mathcal{O}(M_{\star}^{2}R_{0}^{2}) \right]. 
\end{equation} 
We see a strong deviation from the standard inverse square law on scales $\lesssim M_{\star}^{-1}$.  Interestingly, the gravitational force becomes constant on very small scales, which is what one would expect in a $(1+1)$-dimensional theory of gravity.  The effective dimensional reduction of conformal gravity to 2 dimensions \cite{Carlip:2009km} on small scales has been previously noted in \cite{Gegenberg:2017bms}.  Since laboratory tests have confirmed Newton's law on scales $\gtrsim 10^{-8}\,\text{m}$ \cite{Will:2014xja}, this constrains $M_{\star}^{-1} \lesssim 10^{-8}\,\text{m}$.

Equation (\ref{eq:higher order correction}) gives the dependance of the waveform on higher order time derivative initial data; i.e., $\Phi_{2} = (\di^{2}\phi/\di t^{2})_{t=t_{0}}$ and $\Phi_{3} = (\di^{3}\phi/\di t^{3})_{t=t_{0}}$.  It is interesting to re-express this contribution as follows:
\begin{multline}
	\phi_{2}(x) = - \frac{\di}{\di\beta}  \int_{0}^{\beta} dx \iint \frac{\sin\theta\, d\theta \, d\varphi}{4\pi}  \, \frac{x^{2 }J_{1}(\sqrt{\beta^{2}-x^{2}})}{\sqrt{\beta^{2}-x^{2}}}  \left( \frac{\Box\phi}{M_{\star}^{2}}\right)\bigg|_{t=t_{0}} \\ -   \int_{0}^{\beta} dx \iint \frac{\sin\theta\, d\theta \, d\varphi}{4\pi}  \, \frac{x^{2} J_{1}(\sqrt{\beta^{2}-x^{2}})}{\sqrt{\beta^{2}-x^{2}}}  \left( \frac{\di_{t}\Box\phi}{M_{\star}^{3}}\right)\bigg|_{t=t_{0}}.
\end{multline}
Where $\beta = M_{\star}t$.  Now, if we consider plane wave like initial data:
\begin{equation}
	\Phi_{0} = e^{i\mathbf{k}\cdot\mathbf{x}}, \quad \Phi_{1} = (i\omega)e^{i\mathbf{k}\cdot\mathbf{x}},
	\quad \Phi_{2} = (i\omega)^{2}e^{i\mathbf{k}\cdot\mathbf{x}}, \quad \Phi_{3} = (i\omega)^{3}e^{i\mathbf{k}\cdot\mathbf{x}},
\end{equation}	
then we see that $\phi_{2}$ will be negligible compared to $\phi_{1}$ if
\begin{equation}
	 \mathbf{k} \cdot \mathbf{k} \ll M_{\star}^{2}, \quad |\omega| \ll M_{\star}.
\end{equation}
In other words, if the initial data does not involve wavenumbers or frequencies $\gtrsim M_{\star}$, then the contribution of the second and third time derivatives of the initial data to the total waveform will be suppressed:
\begin{equation}\label{eq:approx Kirchhoff}
	\phi(x) \simeq \phi_{F}(x) + \phi_{1}(x).
\end{equation}
Furthermore, if the source does not vary much over scales $\lesssim 1/M_{\star}$, the $\phi_{F}$ will be given by (\ref{eq:source equation limit}), which means that (\ref{eq:approx Kirchhoff}) reproduces the Kirchhoff formula for $\Box \phi = F$.

Finally, we comment on the $\epsilon = -1$ case.  In this situation, it is easy to see that $G_{0}(x-x')$ as given in equation (\ref{eq:conformal Green's}) diverges exponentially as $\sqrt{\tau^{2}-R^{2}} \rightarrow \infty$.  This means that the solutions to (\ref{eq:GW EOM scalar}) are inherently unstable.  This could have easily been guessed from the original wave equation, since any solution $(\Box+M_{\star}^{2})\phi=0$ is both a tachyon and also automatically a solution of $(\Box + M_{\star}^{2})\Box\phi(x) = \alpha F(x)$.  

\section{Discussion}\label{sec:discussion}

In this paper, we have written down the explicit solution for a certain class of fourth order scalar wave equation in terms of an arbitrary source function and initial data involving time derivatives up to order three.  For a certain choice of parameters, the fourth order equation governs the evolution of gravitational waves in conformal gravity.  If the energy scale of the conformal symmetry breaking in such model is of order $M_{\star}$, we find that for initial data and source functions that do not vary much on scales $\lesssim 1/M_{\star}$ our solution reduces down to the regular Kirchhoff formula for the solution of the ordinary massless wave equation.

Do the effects described in this paper have any observational consequences?  Clearly, the answer depends on the characteristic size of a source or initial data compared to $1/M_{\star}$.  If $M_{\star} \gtrsim 10^{-8}\,\text{m}$ as suggested by tests of Newton's law, it is hard to imagine many late universe astrophysical phenomena that would produce gravitational waves with behaviour appreciably different from general relativity.  More promising would be small scale gravitational waves generated in the early universe via inflation, preheating, or some other mechanism.  Investigating the cosmological transfer function between the early and late universe in conformal gravity could be an interesting exercise in the future.

\appendix

\section{Direct calculation of the retarded Green's function}\label{sec:direct}

Here, we present an alternate derivation of the Green's function of $\LL{0}$ using a generalization of the procedure detailed in \S12.1-12.4 of \citet{Poisson:2011nh}.  The PDE to be solved is
\begin{equation}\label{eq:appendix pde}
	\LL{0} G_{0}(x-x') = \delta^{(4)}(x-x').
\end{equation}
The first step involves making the ansatz:
\begin{equation}\label{eq:ansatz 1}
	G_{0}(x,x') = \Theta(t-t') g(\sigma(x,x')),
\end{equation}
where $\Theta$ is the Heaviside function, and
\begin{equation}
	\sigma(x,x') = - \frac{(t-t')^{2} - | \mathbf{x}-\mathbf{x}' |^{2} }{2}
\end{equation}
is Synge's world function, which is a Lorentz invariant.  From (\ref{eq:identities}), we know that
\begin{equation}\label{eq:appendix pde 2}
	\LL{1} G_{0}(x-x') = G_{2}(x-x').
\end{equation}
Let us integrate this equation with respect to $x$ over a bounded spacetime region $\Omega$ containing the point $x'$.  After making use of the divergance theorem, we obtain
\begin{equation}\label{eq:weak form}
	\int_{\di\Omega} \nabla^{\alpha} G_{0}(x-x') d\Sigma_{\alpha} -m_{1}^{2} \int_{\Omega} G_{0}(x-x') d^{4} x = \int_{\Omega} G_{2}(x-x') d^{4} x,
\end{equation}
where $\di\Omega$ is the boundary of $\Omega$, $d\Sigma_{\alpha}$ is the surface element on $\di\Omega$.  Now, let's introduce a coordinate system
\begin{subequations}
\begin{align}
	t & = t' + w\cos\chi \\  x &  = x' + w\sin\chi \sin\Theta \cos\phi,  \\ y & = y' + w\sin\chi \sin\Theta \sin\phi, \\ z & = z' + w\sin\chi \cos\Theta.
\end{align}
\end{subequations}
In these coordinates, $x'$ corresponds to $w=0$ and the metric of flat space is
\begin{equation}
	ds^{2} = -\cos 2\chi \, dw^{2} + 2w\sin 2\chi \, dw \, d\chi + w^{2} [ \cos 2\chi \, d\chi^{2} + \sin^{2}\chi (d\Theta^{2} + \sin^{2}\Theta \, d\phi^{2})].
\end{equation}
We take $\di\Omega$ to be the surface $w = \sqrt{2\epsilon}$ and will ultimately take the $\epsilon \rightarrow 0$ limit.  The only non-vanishing component of the surface element on $\di\Omega$ is
\begin{equation}
	d\Sigma_{w} = w^{3} \sin^{2}\chi \sin\Theta \, d\chi \,  d\Theta \, d\phi.
\end{equation}
In these coordinates, the ansatz (\ref{eq:ansatz 1}) reads
\begin{equation}
	G_{0} = \Theta(w\cos\chi) g(\sigma), \quad \sigma = -\tfrac{1}{2} w^{2} \cos 2\chi
\end{equation}
Inserting this into (\ref{eq:weak form}), we obtain
\begin{align}\label{eq:big integral}
	\int_{\Omega} G_{2}(x-x') d^{4} x = 4 \pi w^{4}  \int_{0}^{\pi}  \sin^{2}\chi \, g'(\sigma)\Theta(w\cos\chi) \, d\chi - m_{1}^{2} \int_{\Omega} G_{0}(x-x') d^{4} x.
\end{align}
Now, let us make an ansatz for $g(\sigma)$ consistent with retarded boundary conditions:
\begin{equation}
	g(\sigma) = \Theta(-\sigma) V(\sigma) + A_{0} \delta(\sigma) + A_{1} \delta'(\sigma) + A_{2} \delta''(\sigma) + \cdots
\end{equation}
Here, $V(\sigma)$ is taken to be a smooth function and the $\{ A_{n} \}_{n=0}^{\infty}$ are constants.  This form immediately implies that the two integrals involving $\delta$-functions derivatives on the righthand side of (\ref{eq:big integral}) vanish.  Changing variables from $\chi$ to $\sigma$, we are left with
\begin{multline}\label{eq:big mess}
	\int_{\Omega} G_{2}(x-x') d^{4} x  = 4\pi\epsilon \int_{-\epsilon}^{0} \sqrt{\frac{\epsilon+\sigma}{\epsilon-\sigma}} V'(\sigma) d\sigma -m_{1}^{2} \int_{\Omega} G_{0}(x-x') d^{4} x - 4\pi \epsilon V(0) \\ - 4\pi A_{0} + \frac{4\pi A_{1}}{\epsilon} - \frac{12\pi A_{2} }{\epsilon^{2}} + \frac{36\pi A_{3}}{\epsilon^{3}} - \cdots
\end{multline}
In the $\epsilon \rightarrow 0$ limit, we can make use of the retarded Green's function solutions (\ref{eq:known Green's}) to show that the integral on the lefthand side vanishes.  Also, the first integral on the righthand side vanishes since $V$ is assumed to be a smooth function.  In order for the limit to exist, we require that the righthand side of (\ref{eq:big mess}) remains finite, which implies
\begin{equation}
	A_{1} = A_{2} = \cdots = 0 \quad \Rightarrow \quad g(\sigma) = \Theta(-\sigma) V(\sigma) + A_{0} \delta(\sigma).
\end{equation}
Putting this form of $g(\sigma)$ into the volume integral of $G_{0}$ over $\Omega$ in (\ref{eq:big mess}) implies that
\begin{equation}
	\lim_{\epsilon \to 0}  \int_{\Omega} G_{0}(x,x') d^{4} x = 0.
\end{equation}
The only surviving term in (\ref{eq:big mess}) yields $A_{0} = 0$.  Hence, we must have
\begin{equation}\label{eq:g non-singular}
	g(\sigma) = \Theta(-\sigma) V(\sigma).
\end{equation}
This is demonstrates that $G_{0}$ is non-singular.

To determine $V$, we consider the PDE (\ref{eq:appendix pde}) in the spacetime region with $t > t'$.  In this region, $x \ne x'$ and (\ref{eq:appendix pde}) become
\begin{equation}
	\LL{1} \LL{2} g(\sigma) = 0.
\end{equation}
It is not difficult to confirm that this equation reduces down to the ordinary differential equation
\begin{equation}
	(2\sigma \di_{\sigma}^{2}+4 \di_{\sigma} - m_{1}^{2})(2\sigma \di_{\sigma}^{2}+4 \di_{\sigma} - m_{2}^{2}) g(\sigma)=0.
\end{equation}
Inserting (\ref{eq:g non-singular}) into this, we obtain
\begin{equation}
	\Theta(-\sigma)[(2\sigma \di_{\sigma}^{2}+4 \di_{\sigma} - m_{1}^{2})(2\sigma \di_{\sigma}^{2}+4 \di_{\sigma} - m_{2}^{2}) V(\sigma)] + 2 \delta(\sigma)[(m_{1}^{2}+m_{2}^{2})V(0)-4V'(0)]=0,
\end{equation}
where we have made use of the distributional identities
\begin{subequations}
\begin{align}
	\Theta'(x) = & \delta(x), \\
	h(x) \delta(x) = & h(0)\delta(x), \\
	h(x) \delta'(x) = & h(0) \delta'(x)-h'(0)\delta(x), \\
	h(x) \delta''(x) = & h(0) \delta''(x)-2 h'(x)\delta'(x) - h''(0)\delta(x), \\
	h(x) \delta'''(x) = & h(0) \delta'''(x) - 3h''(x)\delta'(x)  - 3h'(x)\delta''(x) - h'''(0)\delta(x),
\end{align}
\end{subequations}
The coefficients of $\Theta(-\sigma)$ and $\delta(\sigma)$ in this expression must vanish individually, which leads to an ODE for $V(\sigma)$ plus a boundary condition:
\begin{subequations}
\begin{align}
	\label{eq:V ODE} 0 & = (2\sigma \di_{\sigma}^{2}+4 \di_{\sigma} - m_{1}^{2})(2\sigma \di_{\sigma}^{2}+4 \di_{\sigma} - m_{2}^{2}) V(\sigma),\\ 0 & = (m_{1}^{2}+m_{2}^{2})V(0)-4V'(0). \label{eq:V BC}
\end{align}
\end{subequations}

When $m_{1} \ne m_{2}$, we find the general solution of (\ref{eq:V ODE}) is
\begin{equation}
	V(\sigma) = \frac{a J_{1}(m_{1}\sqrt{-2\sigma}) + b J_{1}(m_{2}\sqrt{-2\sigma}) +  c Y_{1}(m_{1}\sqrt{-2\sigma}) + d Y_{1}(m_{2}\sqrt{-2\sigma})}{\sqrt{-2\sigma}}.
\end{equation} 
Here, $J_{1}$ and $Y_{1}$ are Bessel functions.  Since $V$ is known to be non-singular at $\sigma = 0$, we set $c=d=0$.  Imposition of the boundary condition (\ref{eq:V BC}) then yields a relation between $a$ and $b$:
\begin{equation}
	m_{2}a + m_{1}b = 0.
\end{equation}
Finally, direct substitution into (\ref{eq:appendix pde 2}) fixes $a$ and $b$.  We arrive at the final answer:
\begin{align}
	G_{0}(x-x')  = \frac{\Theta(t-t')\Theta(-\sigma)}{4\pi\sqrt{-2\sigma}} \left[ \frac{ m_{1} J_{1}(m_{1}\sqrt{-2\sigma}) - m_{2} J_{1}(m_{2}\sqrt{-2\sigma}) }{m_{1}^{2} -m_{2}^{2}} \right].
\end{align}
In agreement with the calculation of Section \ref{sec:Green's}.

A very similar analysis of the degenerate $m_{1} = m_{2} = m$ case leads to the retarded Green's function
\begin{align}
	G_{0}(x-x')  = \frac{\Theta(t-t')\Theta(-\sigma)J_{0}(m\sqrt{-2\sigma})}{8\pi}.
\end{align}

\acknowledgments

I would like to thank Jack Gegenberg for early collaboration on some of the results presented in this paper, and NSERC of Canada for financial support.

\bibliographystyle{apsrev4-1}
\bibliography{GWs_YM}

\end{document}